\documentclass[runningheads]{llncs}
\usepackage[T1]{fontenc}
\usepackage{graphicx}
\usepackage[export]{adjustbox}
\usepackage{float}
\usepackage{placeins}
\usepackage{longtable}   
\usepackage{multirow}
\usepackage{tabularx}
\usepackage{lscape}
\usepackage{booktabs}
\usepackage{caption}
\usepackage{xcolor}
\usepackage{makecell}
\usepackage{comment}
\usepackage{array} 
\begin{document}

\title{Challenges and Contributions in Quality of AI-Based Software: A Systematic Mapping Study}

\titlerunning{Challenges and Contributions in Quality of AI-Based Software}
%
%

\author{ Maryum Hamdani\inst{1}\thanks{Corresponding author.} \and Mateen Ahmed Abbasi\inst{2} \and Marko Jäntti\inst{1} \and Markku Tukiainen\inst{1} }

\authorrunning{M. Hamdani et al.}

\institute{
Department of Computer Science, University of Eastern Finland (UEF), Finland\\
\email{\{maryum.hamdani,marko.jantti,markku.tukiainen\}@uef.fi}
\and
Faculty of Information Technology, University of Jyväskylä, Jyväskylä, Finland\\
\email{mateen.a.abbasi@jyu.fi}
}

\maketitle              
\begin{abstract}

Artificial Intelligence (AI) is increasingly embedded in modern software systems, raising important questions about how its quality should be defined, assessed, and assured. This paper presents a Systematic Mapping Study (SMS) on the quality of AI-based software. The study synthesizes primary studies published between January 2020 and January 2026 and selected from five electronic data sources. A total of 33 primary studies were included after automated search, screening, and snowballing. The results identify six recurring challenge categories, with the most prominent being limitations in existing quality assessment models, followed by issues in non-functional requirement management, quality-aware development, and quality assurance. The findings suggest a call for collaboration of researchers and industrial practitioners with standardization organizations, that could possibly devise comprehensive quality assessments and their measurement methods.


\end{abstract}
\section{Introduction}
Artificial Intelligence (AI)-based software refers to software systems that have AI components as an integral part of their functionality \cite{a31}. These AI components are predominantly implemented by Machine learning (ML) and Deep Learning (DL) approaches  \cite{a45}. As AI enabled functionality becomes increasingly embedded in production software systems, concerns about software quality have become more difficult and more important to address. \par

Ensuring quality in AI-based software is challenging because such software systems are often data dependent, probabilistic, and context sensitive. Unlike conventional deterministic software, their behavior varies due to any change in input data, operating condition, and user interaction which complicates, specification, verification, and evaluation \cite{a10}\cite{a38}. Concerns regarding the quality of AI-based software has been amplified by the increasing number of publicly reported AI incidents. According to HAI index 2025, the database recorded 233 AI-related incidents in 2024, which is a 56.4\% increase over 2023, indicating the growing consequences of AI failures and misuse \cite{StanfordAIIndex2025}. \par
Researchers have explored multiple dimensions of quality in AI-based software, investigating AI-specific quality attributes \cite{a1}, evaluation methods \cite{a4}, and quality frameworks \cite{a14}. Likewise, secondary studies such as systematic reviews \cite{a32}, mapping studies \cite{a39}, and taxonomies \cite{a40} have examined selected aspects of the area. However, the evidence remains fragmented due to the update in research explorations, across subtopics, terminologies and software engineering concerns.

This motivates a systematic mapping study to synthesize the state of the art in quality in AI-based software. The objective of this study is to identify the major challenges reported in relation to quality of AI-based software and the contributions proposed to address those challenges and characterize the maturity of evidence base in terms of research type, contribution type and research methods.  To achieve this objective, we used the Systematic Mapping Study (SMS) guidelines as a research method. The mapping covers primary studies published from 1 January 2020 to 15 January 2026, a period of rapid development in AI-based software.

Although prior secondary studies have examined quality models, software quality concerns, and quality assessment taxonomies for AI-based software, the existing evidence remains dispersed across different analytical scopes. Some studies focus primarily on quality models, while others emphasize quality assessment or selected quality attributes. In contrast, this SMS provides an integrated view of the area by jointly mapping: (i) the challenges reported in relation to the quality of AI-based software, (ii) the contributions proposed to address those challenges, and (iii) the maturity of the evidence base in terms of research type, contribution type, and research method. This combination enables a broader understanding not only of what has been studied but also of how mature, validated, and practically actionable the existing evidence is.

Moreover, this SMS covers the state of evidence until January 2026, which is the time when fast advancements were achieved due to increasing application of AI in industry, introduction of AI regulations, and publication of new AI-specific quality standards like ISO/IEC 25059:2023. Therefore, the novelty of this SMS can also be seen in its integrative analysis of challenges, solutions, standardization gaps, and empirical evidence maturity. The remainder of the paper is structured as follows. Section 2 describes the research methodology. Section 3 presents the results. Section 4 discusses the findings and implications. Section 5 outlines threats to validity. Section 6 concludes the paper.


\section{Research Methodology}
The research problem of the study is: What is the state of the art with respect to quality of AI-based software in current literature? Based on the research problem, we formulated three research questions: RQ1.) What are the challenges reported regarding the quality of AI-based software? RQ2.) What are the contributions reported to address the challenges regarding the quality of AI-based software? RQ3.) What is the classification of selected studies in terms of contribution types, research methods, and research types? We used the guidelines by Peterson et al.\cite{a48}\cite{a42} to conduct a systematic mapping study.
\subsection{Search Process}
The steps followed to conduct this SMS are illustrated in Fig. 1. As shown, the search process comprises of two phases. In phase 1, we conducted the primary search by following the steps outlined below.
\begin{figure}[ht]
\centering
\includegraphics[width=0.9\textwidth, trim={0cm 200cm 0cm 0cm}, clip]{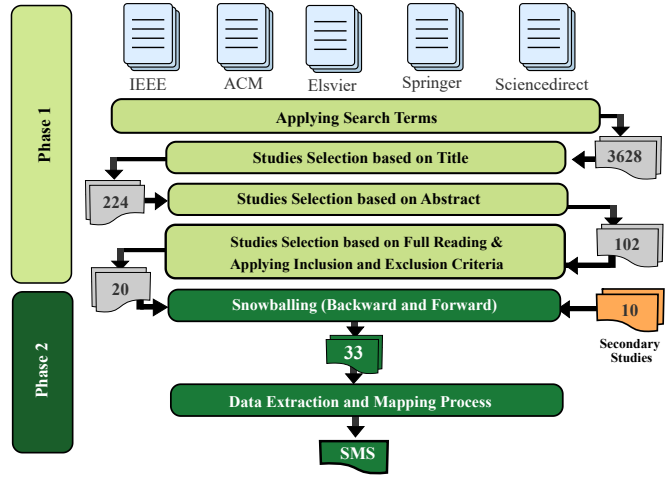}
    \caption{Study selection workflow}
    \label{fig:research-type}
\end{figure}

    \textbf{\textit{Search terms selection:}} 
    We defined the search terms in Table 1 using the PICO (Population, Intervention, Comparison and Outcomes) framework as per guidelines of Kitchenham et al. \cite{a41}. In order to ensure that search string captures maximum number of related studies, we performed multiple iterations of pilot searches, hence improved the search string.

\begingroup
\footnotesize
\begin{table}[ht]
\caption{Search string using the PICO framework}
\label{tab:placeholder}
\centering
{\footnotesize
\setlength{\tabcolsep}{3pt}
\renewcommand{\arraystretch}{0.95}
\begin{tabular}{@{}p{0.15\linewidth}p{0.78\linewidth}@{}}
\hline
\textbf{Element in PICO} & \textbf{Search string part} \\ \hline

Population &
``Artificial intelligence software'' OR 
``AI software'' OR 
``AI-based software'' OR 
``Machine learning software'' OR 
``ML software'' OR 
``Artificial intelligence enabled software'' OR 
``AI-enabled software'' \\ \hline

Intervention &
``Quality assessment'' OR 
``Quality evaluation'' OR 
``Quality assurance'' OR 
``Quality model'' OR 
``Quality attribute'' OR 
``Quality characteristic'' OR 
``Nonfunctional requirement'' OR 
``Non-functional requirement'' \\ \hline

Comparison & Not applicable \\ \hline

Outcome & Metric OR measurement OR method \\ \hline
\end{tabular}
}
\end{table}
\endgroup
    \textbf{ \textit{Title abstract and full text screening:}} 
    {We conducted the study in two phases: automated database search and manual snowballing. The automated search covered primary studies published from 1 January 2020 to 15 January 2026 and was executed across Electronic Data Sources (EDS) including IEEE Xplore, ACM Digital Library, SpringerLink, ScienceDirect, and Wiley Online Library. The search returned 3,628 records (IEEE Xplore: 873, ACM Digital Library: 899, SpringerLink: 616, ScienceDirect: 888, Wiley Online Library: 352). The selection of EDS is based on the guidelines by Zhang et al \cite{a43}. After title-based screening and duplicate removal, 224 records remained for abstract screening. Abstract screening retained 102 studies for full-text assessment. Full-text screening against the inclusion and exclusion criteria resulted in 20 primary studies. In the second phase, backward and forward snowballing identified 13 additional studies (5 backward and 8 forward), resulting in a final corpus of 33 primary studies.} \par

{The full-text assessment was conducted by the first and second authors using the complete inclusion and exclusion criteria presented in Table 2. 
To improve the reliability of the study selection process, the first and second authors first conducted a calibration round on a subset of candidate studies. The inclusion and exclusion criteria for the articles were discussed and revised in order to reduce any potential misinterpretation. After completing the calibration process, both authors independently assessed the candidate studies during the full-text screening phase. Disagreements were resolved through discussion, with the third author being included when necessary. This procedure was used to reduce individual selection bias and to ensure consistent application of the selection criteria.


\begin{longtable}{@{}p{0.48\linewidth}p{0.50\linewidth}@{}}
\caption{Inclusion and exclusion criteria}
\label{tab:my-table}\\
\hline
\textbf{Inclusion criteria} & \textbf{Exclusion criteria} \\ \hline
\endfirsthead

\hline
\textbf{Inclusion criteria} & \textbf{Exclusion criteria} \\ \hline
\endhead

\hline
\endfoot

\hline
\endlastfoot

IC1. Include primary studies that address quality aspects of AI/ML-based software systems.
&
EC1. Exclude studies that address quality aspects of traditional software systems, i.e., without an AI component.
\\ \hline

IC2. Include studies published from 2020 onwards.
&
EC2. Exclude studies on AI/ML-based software that do not address quality-related concerns.
\\ \hline

IC3. Include studies from the software engineering domain only.
&
EC3. Exclude studies published before 2020.
\\ \hline

IC4. Include studies written in English only.
&
EC4. Exclude studies that solely address data quality aspects.
\\ \hline

IC5. Include primary studies published in journals, conferences, surveys, workshops, and book chapters.
&
\multirow{2}{=}{\parbox[t]{\linewidth}{EC5. Exclude duplicate studies, conference/workshop papers extended into journal articles, doctoral consortium papers, dissertations, secondary studies, posters, and editorials.}}
\\ \cline{1-1}

IC6. Include studies with four pages or more.
&
\\ \hline

\end{longtable}

\subsection{Data Extraction and Mapping Process}
Table 3 summarizes the data extraction schema used in this study. Data items D1 to D5 present the demographics of the final selected studies. We performed inductive thematic analysis \cite{a47} across the selected studies to extract D6 and D7. This analysis systematically identified and categorized: the primary research challenge that each study reported in relation to quality of AI-based software (D6), and the proposed contributions each study reported to address the corresponding challenge (D7). Furthermore, the selected studies were classified by research type (D8) and contribution type (D9), following the classification schemes proposed by Wieringa et al. \cite{a49} and Petersen et al.\cite{a48}. Lastly, we extracted primary research methods (D10), employed in each study to produce its results based on the taxonomy of software engineering research methods outlined by Glass et al. \cite{a54}. 
The replication package is available at \cite{anonymous_2026_19545171} as anonymous and contains the search strings, screening decisions, data extraction sheet and study classifications.

{\small{
\begin{longtable}[c]{l l p{0.4\textwidth} c}
\caption{Data extraction schema}
\label{tab:my-table}\\
\hline
Code & Data Item         & Description                                                      & RQ                            \\ \hline
\endfirsthead
\endhead
D1   & Index             & The ID of the   study                                            & \multirow{5}{*}{Demographics} \\ \cline{1-3}
D2   & Title             & The title of the study                                           &                               \\ \cline{1-3}
D3   & Year              & The publication year of study                                    &                               \\ \cline{1-3}
D4   & Publisher         & The publisher of the study                                       &                               \\ \cline{1-3}
D5   & Publication Type  & Journal, Conference, Workshop, Symposiums                        &                               \\ \hline
D6 &
  Challenges &
  The main challenge that each study reported   in relation to the quality of AI-based software, extracted by guidelines of   {}\cite{a47} &
  RQ1 \\ \hline
D7 &
  Contributions &
  The key contributions each study proposed   in addressing challenge regarding quality of AI-based software extracted by   guidelines of \cite{a47} &
  RQ2 \\ \hline
D8 &
  Research type &
  The research type classification proposed   by Wieringa et al. {}\cite{a49} &
  \multirow{3}{*}{RQ3} \\ \cline{1-3}
D9   & Contribution type & We classified the contribution types as per   \cite{a48}              &                               \\ \cline{1-3}
D10  & Research methods  & The research methods reported in the selected   studies \cite{a54} &                               \\ \hline
\end{longtable}}}
\section{Results}
\subsection{RQ.1) What are the challenges reported regarding the quality of AI-based software?
}
{We identified 6 categories of recurring challenge in the primary studies. Table 4 summarizes the categories, representative evidence, and associated study IDs; the description of challenge below highlights only the main patterns.
\begingroup
\footnotesize
{\footnotesize
\setlength{\tabcolsep}{3pt} 
\renewcommand{\arraystretch}{1.0}
\begin{longtable}{@{}
p{0.20\linewidth}
p{0.72\linewidth}
p{0.06\linewidth}
@{}}
\caption{Challenges reported in the primary studies}
\label{tab:my-table}\\
\hline
Category (n) & Challenge & Study ID \\ \hline
\endfirsthead

\hline
Category (n) & Challenge & Study ID \\ \hline
\endhead

\hline
\endfoot

\hline
\endlastfoot

1. Limitations in existing quality assessment models (13)
& Additional AI-specific sub-characteristics in the extended model, i.e., ISO/IEC 25059, are too limited.
& S1 \\ \cline{2-3}

& ISO/IEC 25059 is high level and incomplete; it is necessary to define a set of metrics and thresholds.
& S5 \\ \cline{2-3}

& The key characteristics proposed in ISO/IEC 25059:2023 do not yet have specific testing methods for their implementation.
& S10 \\ \cline{2-3}

& ISO/IEC 25059 has some gaps when it comes to the coverage of Act requirements.
& S25 \\ \cline{2-3}

& The quality-in-use model on which current AI quality management standards (ISO/IEC 25059) are based does not take into account various stakeholders.
& S32 \\ \cline{2-3}

& \ldots & \\ \hline

2. Non-Functional Requirement (NFR) management issues (6)
& Lack of detailed insights on quality trade-offs observed in industrial practice and how companies address them.
& S7 \\ \cline{2-3}

& Managing NFRs is particularly challenging due to the differing nature and definitions of NFRs for ML systems, including non-deterministic behavior.
& S3 \\ \cline{2-3}

& Despite the importance of NFRs in ensuring the quality of ML systems, our understanding of these aspects (definition, measurement, scope, and comparative importance) is lacking.
& S16 \\ \cline{2-3}

& \ldots & \\ \hline

3. Lack of quality-aware development approaches (6)
& There is limited literature focused on discovering and understanding quality issues encountered by practitioners when building MLSSs.
& S2 \\ \cline{2-3}

& Solutions have been proposed to automate the development of ML systems. However, an approach that takes into account the new quality concerns needed by ML systems is still missing.
& S8 \\ \cline{2-3}

& Engineering practices for building these systems remain poorly understood compared to those for conventional software systems.
& S33 \\ \cline{2-3}

& \ldots & \\ \hline

4. Inadequate QA for AI-based software (5)
& There is a lack of standardized approaches for Quality Assurance (QA) of AI-based systems.
& S18 \\ \cline{2-3}

& AI-enabled systems require attention to software quality assurance in general and code quality in particular.
& S20 \\ \cline{2-3}

& Due to the statistical nature of machine learning, traditional quality assurance approaches are often insufficient.
& S29 \\ \cline{2-3}

& \ldots & \\ \hline

5. Limitations in existing QMF (1)
& The Quality Management Framework (QMF) of AI systems is fragmented and incomplete.
& S27 \\ \hline

6. Cross domain (2)
& No one has systematically investigated the impact of using bindings for ML frameworks on ML software quality.
& S6 \\ \cline{2-3}

& \ldots & \\ \hline
\end{longtable}
}
\begin{center}
\end{center}
\endgroup
\textbf{Category 1 Limitations in existing quality assessment models. }This category of challenge includes the studies that highlighted the limitations of existing quality models for the assessment of AI-based software. As an example, Ramos et al. \cite{a1}  states that the extended quality attributes in ISO/IEC 25059: 2023 \cite{iso25059}, to address unique characteristics of AI-based software, are too limited. Similarly, Oviedo et al. \cite{a4}  reflects on the absence of metrics in ISO/IEC 25059: 2023 \cite{iso25059} to measure the quality. Moreover, De Sanctis et al. in \cite{a5}, mentioned that the updated ISO 25010:2023 \cite{iso25010:2023} fails to consider stakeholder needs of modern systems. \par
\textbf{Category 2 NFR management issues}. This category contains primary studies that report issues in NFR management for ML-enabled Software. NFR management here encompasses specification, definition, scoping, and prioritization of NFR. Since these software systems exhibit non-deterministic nature therefore, existing methods and techniques of NFR management have become inapplicable \cite{a16}. Indykov et al. in \cite{a7} and Habibullah et al. \cite{a16}, investigates the challenges practitioners encounter in 1) prioritizing NFR types for ML-enabled Software and 2) establishing appropriate measurement criteria in industry. Habib et al. \cite{a3} highlighted the need to scope NFR over different system components (e.g. data, models, and code). Whereas Heck et al. \cite{a9} focused on how quality requirements for AI monitoring platform can be defined using quality model.\par
\textbf{Category 3 Lack of quality-aware AI-based software development approaches}. All primary studies that indicates the absence of quality-aware approaches while developing AI-based software are classified in this category. Côté et al. \cite{a2}  reported that actual issues encountered by practitioners while developing ML-based software systems are not well investigated. D’Alessio et al. \cite{a8} observed that despite the prevalence of automated development solutions for ML-based software systems, the systematic approaches for integrating quality concerns—such as privacy, fairness, and reliability—into their development process is notably lacking. Similarly, Bucaioni et al. \cite{a55} further asserted that understanding of engineering practices for developing ML-intensive software systems lag significantly compared to those for conventional software systems. \par
C\textbf{ategory 4 Inadequate QA approaches.} Primary studies that highlights inadequate quality assurance practices for AI-based software are accumulated in this category. Felderer et al. \cite{a18} stated that there is a lack of agreement regarding the approaches available for implementing quality assurance in AI-based systems, leading to conflicting solutions among practitioners. Golendukhina et al. \cite{a20} also emphasized, a critical need for enhanced software quality assurance (SQA) practices in AI-enabled systems, with particular emphasis on maintaining code quality standards. Similarly, Poth et al. \cite{a28} stated that adequate quality measurement and assurance techniques for ML-based software systems are still absent. \par
\textbf{Category 5 Limitation in existing QMF.} This category captures organization level limitations in existing quality management frameworks. Although represented by a single study in the current corpus, it highlights a distinct objective. Santhanam \cite{a27} states the traditional quality management process (consisting of i) requirement management, ii) defect management, iii) change management, iv) test management, v) DevOps processes, vi) operations management, and vii) project management) is inadequate when applied to ML-based software. As ML-based software reflects a different nature, current quality management framework seems inapplicable for building software with different quality attributes such as fairness, biasness etc. \par
\textbf{Category 6 Cross-domain. }
This category groups studies that address relevant quality issues not captured by the above categories, for instance, Li et al. \cite{a6} uncovered that the impact of ML framework bindings on AI-based software quality is systematically unexplored. In contrast, Brower-Sinning et al. in \cite{a12} reported about the lack of information received at developer’s end regarding the design decisions, due to which they consider limited properties for ML model testing, resulting in failure in model's deployment and operation. \par
 It is important to mention here that a few studies fulfill criteria to be classified under multiple categories simultaneously. For example, the \cite{a7} can be classified into the category of “Lack of quality-aware practices” and “NFR  management issues”. In such cases, we classified the study according to its primary focus.
\subsection{RQ.2) What are the contributions reported to address the challenges regarding the quality of AI-based software?
}
We identified 11 contribution categories from the primary studies (see Table 5). These categories reflect a diverse set of solutions reported to address the identified challenges (RQ1). Each category represents a recurring type of solution reported across selected studies. In table 5, the rows are grouped by challenge category (see Table 4) it targets. Additionally, it contains one or more contribution. The third column presents a focused description of the solution and finally are the serial number of studies (S1-S33). All categories are briefly described below.\par

\begingroup
\footnotesize
{
\setlength{\tabcolsep}{3pt}
\renewcommand{\arraystretch}{0.95}
\setlength{\LTleft}{0pt}
\setlength{\LTright}{0pt}

\begin{longtable}{@{}p{0.05\linewidth}p{0.25\linewidth}p{0.50\linewidth}p{0.14\linewidth}@{}}
\caption{Contributions addressing the identified challenges}
\label{tab:contributions}\\
\hline
\textbf{No} & \textbf{Contribution category} & \textbf{Contribution focus} & \textbf{Study ID} \\
\hline
\endfirsthead

\hline
\textbf{No} & \textbf{Contribution category} & \textbf{Contribution focus} & \textbf{Study ID} \\
\hline
\endhead

\hline
\endfoot

\hline
\endlastfoot

\multicolumn{4}{@{}l@{}}{\textbf{Category 1: Limitations in existing quality evaluation models}} \\
\hline
1. & Quality model enhancement & Extending and improving existing quality evaluation models for AI-based software & S1, S4, S10, S19, S22, S24 \\
2. & Quality model development & Creating novel quality evaluation models for AI-based software & S13, S14, S21, S23, S32 \\
3. & Conceptual redefinition & Redefining fundamental quality concepts & S5, S25 \\
\hline

\multicolumn{4}{@{}l@{}}{\textbf{Category 2: NFR management issues}} \\
\hline
1. & Approaches for NFR scoping, definition, and prioritization & Scoping, defining, and prioritizing non-functional requirements & S3, S7, S11, S16 \\
2. & Quality model-based NFR specification & Using quality models for NFR specification & S9, S15 \\
\hline

\multicolumn{4}{@{}l@{}}{\textbf{Category 3: Lack of quality-aware development approaches}} \\
\hline
1. & Quality-aware development support & Tools and methods for quality-driven development of AI-based software & S8, S26, S30, S33 \\
2. & Quality-aware practices characterization & Empirical analysis of quality challenges and practices for AI-based software & S2, S17 \\
\hline

\multicolumn{4}{@{}l@{}}{\textbf{Category 4: Inadequate quality assurance for AI-based software}} \\
\hline
1. & QA methodological frameworks & Comprehensive quality assurance process frameworks & S28, S29, S31 \\
2. & QA challenges and strategies characterization & Identifying and addressing QA obstacles & S18, S20 \\
\hline

\multicolumn{4}{@{}l@{}}{\textbf{Category 5: Limitations in existing QMF}} \\
\hline
1. & Preliminary QMF framework for AI-based software & Preliminary Quality Management Framework & S27 \\
\hline

\multicolumn{4}{@{}l@{}}{\textbf{Category 6: Cross-domain}} \\
\hline
1. & Impact on quality via framework binding & Investigation of the impact on ML software quality (correctness and time cost) using bindings for ML frameworks & S6 \\
2. & Quality attribute-driven ML model testing & Approach to generate ML test cases based on quality attributes & S12 \\
\hline

\end{longtable}
}
\endgroup
\textbf{Category 1.1 Quality model enhancement.} Studies in this category aim to enhance existing quality assessment models. For instance, Ramos et al. \cite{a1} proposed an extension to quality models in ISO/IEC 25059:2023 \cite{iso25059}, based on a survey conducted from practitioners. They extended its product quality model by adding a quality dimension “Humanity” and 30+ sub-attributes and added 7 sub-attributes to the quality-in-use models . Similarly, Kelly et al. \cite{a22} presented an extension to the ISO/IEC 25059:2023 \cite{iso25010:2023} product quality model for safety-critical systems. The enhanced model includes three additional quality dimensions and thirty-five sub-attributes related to safety and data quality. Guo et al. \cite{a10} extended the ISO/IEC 25059:2023\cite{iso25059} by presenting metrics and measurement methods for AI specific quality attributes which differentiates ISO/IEC 25059:2023 \cite{iso25059} from ISO/IEC 25010:2011 \cite{iso25010}.\par
\textbf{Category 1.2: Quality model development. }This category deals with primary studies that aim to address “limitations in existing quality assessment model” by proposing a novel quality model. Dell’Anna et al. \cite{a21}, proposed a quality model for Human-AI system from teaming perspective. Likewise, Siebert et al. \cite{a13}, developed a systematic process and meta-model for constructing quality models for ML-based system and validated with an industrial case study.\par
\textbf{Category 1.3: Conceptual redefinition.} Primary studies in this category argue with help of evidence and claims for redefining the notion of quality for AI-based software. De Sanctis et al. \cite{a5} emphasized the need of re-examining quality definitions in quality model provided by ISO/IEC 25010:2023 \cite{iso25010:2023} to reflect evolving stakeholder needs, such as i) societal and environmental well-being, ii) diversity, non-discrimination and fairness, along with iii) human agency and oversight, human autonomy. \par
\textbf{Category 2.1: Approaches for NFR management.} This category comprises of primary studies that present approaches for defining, scoping, prioritizing NFRs for ML-based systems. For instance, Habibullah et al. \cite{a3} addressed the NFR management issue by introducing a structured framework for defining, scoping, and quantifying critical NFRs important to the ML system under development. Correspondingly, Habibullah et al. \cite{a16} empirically explored the approaches that practitioners use in industry for defining and measuring NFRs over the multiple aspects of ML-system including the model component, data, and system as a whole.\par
\textbf{Category 2.2: Quality model-based NFR specification.} This category encompasses the primary studies that employ the quality model as a standard dictionary for the specification of quality requirements. Heck et al. \cite{a9} utilized Heck’s quality model \cite{a53}, as a reference framework to systematically define quality requirements for AI Wildflower Monitoring Platform. While Haindl et al. \cite{a15} specified quality requirements for human-AI teaming platform using ISO 25010:2011 \cite{iso25010}. \par
\textbf{Category 3.1: Quality-aware development support.} In this category, the primary studies develop tools and methods that integrate quality considerations throughout the development lifecycle of AI-based software. Sikand et al. \cite{a26} presented a questionnaire-based methodology which enables the practitioner to evaluate the overall sustainability throughout development process. Similarly, D'Aloisio et al. \cite{a8} developed a low-code platform, which uses Extended Feature Models (ExtFM) within a Software Product Line (SPL) to select and combine the features for satisfying specific quality constraints.\par
\textbf{Category 3.2: Quality-aware practices characterization.} This category accumulates all primary studies that investigate the strategies employed by industrial practitioners to mitigate challenges while building AI-based software. Côté et al. \cite{a2} investigated quality issues encountered by thirty-seven practitioners and developed an empirically validated catalogue, which covers causes, consequences, and mitigation strategies for each issue encountered in development of Machine Learning Software System (MLSS). Also, Valentina et al. \cite{a17} reported common quality issues and corresponding solutions in developing AI enabled system based on the experience of interviewing three research groups. Moreover, the author emphasizes on training of developers for reducing technical debt in production of AI enabled systems. \par
\textbf{Category 4.1: QA methodological framework.} This category contains the primary studies that contribute methodological frameworks for systematically assuring quality for ML-based software. Poth et al. \cite{a28} proposed EvAIa, a questionnaire-based approach that provides recommendations to ensure the capabilities AI models and making limitations transparent. Likewise, Baresi et al. \cite{a56} developed a conceptual framework consisting of LLM-based agents known as Artificial Doppelgänger Agents (ADAs). These ADAs intend to evaluate the socio-critical system of pre-production and runtime QA iteratively and ensure its compliance with societal expectations.\par
\textbf{Category 4.2: Characterization of QA challenges and strategies.} Primary studies in this category comprises of empirical and analytical research that characterize QA challenges and document practitioner strategies for assuring quality in AI-based systems. Felderer et al. \cite{a18} identified eight major QA challenges for AI-based systems and established foundational terminology, emphasizing the necessity for cross-disciplinary collaboration between software engineering and AI communities. Golendukhina et al. \cite{a20} investigated the SQA strategies adopted by practitioners , through an exploratory interview study with ten Austrian companies developing AI-enabled systems,. They revealed 11 categories of strategies adapted by Small and Medium Enterprise (SME) in the development, integration, and maintenance of AI/ML components.\par
\textbf{Category 5.1: Preliminary Quality Management Framework (QMF) for AI-based software. }This category addresses limitations in existing quality management frameworks for AI-based systems. P. Santhanam et al. \cite{a27} proposed initial quality management framework AI/ML based applications. This framework means to address the challenges of incorporating AI components into business-critical application. \par
\textbf{Category 6 Cross-domain.} This category contains contributions of primary studies that address the distinctive quality challenges in AI-based software that have not been investigated in any other study within this mapping.   Li et al. \cite{a6}  investigates impact of ML software quality (correctness and time cost) using bindings for ML frameworks for DL model training and inference. Brower-Sinning et al. in \cite{a12} presented an approach based on quality attribute (QA) scenarios to elicit and define system and model-relevant test cases. \par

\subsection{RQ.3) What are the distribution and characteristics of selected studies in terms of research types, contribution types, and research method?}
Fig.2 shows the distribution of all selected studies by research type according to the classification proposed by Wieringa et al \cite{a49}. Ten studies  S2, S7, S15, S16, S19, S20, S26, S28, S29, and S33 are classified as evaluation research, i.e., investigate problem in practice. In contrast, eight studies S3, S4, S8, S9, S10, S11, S21, S22 present solution proposals without full-scale validation. Six studies  S12, S17, S18, S24, S25, S30 report personal experiences from real-life projects. \par
 \begin{figure}[ht]
\centering
\includegraphics[width=.96\textwidth]{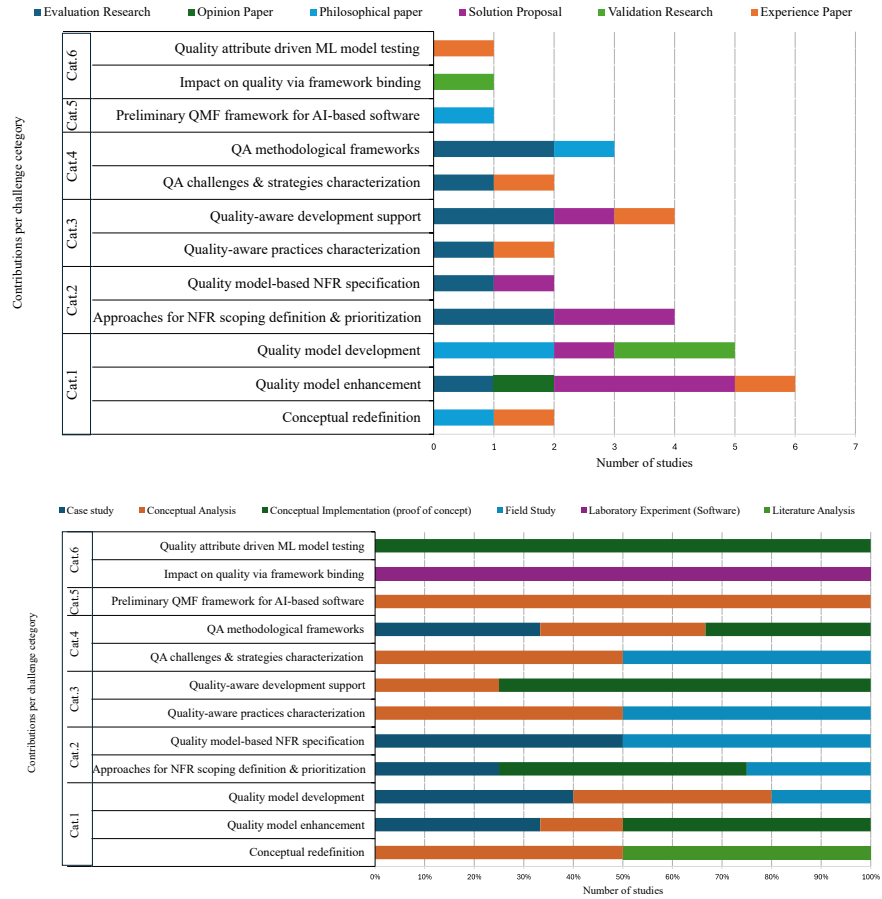}
\caption{Research type and method classification by challenge and contribution category}
\label{fig:research-type}
\end{figure}

\begingroup
\footnotesize
\begin{table}[ht]
\centering
\caption{Distribution of contribution types across the selected studies}
\label{tab:placeholder}
{\footnotesize
\setlength{\tabcolsep}{3pt}
\renewcommand{\arraystretch}{0.95}
\begin{tabular}{@{}p{0.16\linewidth}p{0.48\linewidth}p{0.31\linewidth}@{}}
\hline
\textbf{Contribution type} & \textbf{Description} & \textbf{Studies} \\ \hline

Model &
A conceptual model proposed by a selected study, or an extension of an existing model.
&
S1, S13, S14, S21--23, S27, S32
\\ \hline

Method &
Studies that propose guidelines or methodologies for evaluating quality.
&
S3, S4, S8, S9, S10, S11--13, S19, S22, S24, S29, S31
\\ \hline

Tool &
An automated software tool or checklist to support quality evaluation.
&
S4, S32
\\ \hline

Open item &
Challenges or strategies reported by the selected studies.
&
S2, S5, S6, S7, S15--18, S20, S25, S28, S30
\\ \hline

Metric &
A granular measurable item used to evaluate the quality of AI-based software.
&
S4, S10, S15, S19, S26
\\ \hline

\end{tabular}
}
\end{table}
\endgroup

Furthermore, we categorized five studies S5, S23, S27, S31, S32, as philosophical research, since they offer a conceptual solution to the underlying problem, without the detailed characteristic of a formal solution. Three studies S6, S13, S14 are classified as validation research, since they evaluate proposed solutions in a controlled or experimental setting that has not yet been applied in practice. Finally, one study (S1) is categorized as an opinion paper, as it presents the author's personal perspective without empirical evidence. \par

Table 6 presents distribution of selected studies from contribution types as per classification defined in \cite{a48}. Note: A number of selected studies have produced more than one contribution. The most common contribution type are  guidelines and methods for evaluating software quality. In contrast, the tool is the least populated contribution proposed by the selected studies.\par
Fig. 2 shows the distribution of selected studies with respect to the research methods\cite{a54}. Ten studies studies S3, S4, S8, S10, S11, S12, S22, S26, S28, and S33 have applied conceptual implementation, that aimed to demonstrate the proof of concept. Following that nine studies  S1, S17, S18, S23, S25, S27, S30, S31, and S32 employed conceptual analysis- a research method that does not provide realistic evaluation method. Seven studies S7, S9, S10, S13, S14, S19, S24, and S29 used the research method case studies offering realism to a certain level, e.g (industrial setting). Five studies  S2, S15, S16, S20, and S21 have conducted a field study to investigate the quality practices in real world. One study S6, has used laboratory experiment with software, and one study performed analysis from literature  S5. It is important to mention here that a number of studies have used multiple research methods, however it is classified based on the strongest evidence using one for the results section, the replication package \cite{anonymous_2026_19545171} shows the secondary research methods used by the studies. Examples include S26, and S29 conducted a pilot and S33 conducted a workshop further to validate the results that can be classified as laboratory experiment (using humans). \par

\vspace{-6pt}
{\footnotesize

\section{Discussion}
This study mapped the research landscape regarding the quality of AI-based software based on three research questions. The results show that the reported challenges span requirements engineering, development practices, quality assessment,  quality assurance and quality management as summarized in table 4.  The most prominent challenge emerged  is\textit{ Limitations in existing quality assessment models} covering 40 \% of studies i.e.(13 out of 33) . The dominance of this category indicates that current quality assessment models provided by international standards are insufficient to capture distinctive characteristics of AI-based software systems. One of the reason is rapid evolution of AI and its expansion across multiple industries, which unveils the new characteristics and impact on stakeholders. Thus, one key take-away is that  quality models provided by international standards are considered important in research, for assessing and determining the quality of AI-based software. However, these models insufficient to cope up with fast pace of AI evolution.  \par
According to the findings of RQ2, research efforts have been predominantly directed towards \textit{quality model enhancement} (6 out of 33), followed by the second most populated category \textit{quality model development} ( 5 out of 33). This reflects  continuous efforts by research community in both refining and constructing new quality models for AI-based software. However, the classification results from RQ3, show that many of these contributions remains non empirical or early stage. This indicates that research on quality models for AI-based software remains largely conceptual and theoretical in nature.   Furthermore, the case studies identified in these studies are predominantly conducted in simulated environments rather than industrial settings, which limits their practical generalizability. This highlights a significant gap in the literature, causing the need for rigorous empirical studies conducted in real-world industrial contexts to validate existing quality model extensions and new quality models developments.\par 
Another interesting finding drawn from cross-facet analysis reveals evaluation research  and field study are comparatively more dominant in studies addressing NFR management than in any other identified category. This suggests that among all categories, NFR management stands out as the most empirically grounded area, where researchers investigated practices from industrial settings. Additionally, this study  revealed a significant gap in assessment tools for the quality of AI-based software, i.e., contributed by (2 out of 33) studies \par
Taken together the findings suggests a call for the collaboration of researchers and industrial practitioners with standardization organizations that could possibly devise comprehensive quality assessment models and their measurement methods. Therefore, future research should focus on operationalizing AI specific quality attributes, strengthening empirical validation, and integrating quality concerns more systematically.
\section{Threats to validity}
We addressed threats related to search strategy bias by conducting pilot searches, which helped us identify a broader set of relevant synonyms and refine the search string. Furthermore, the final search string was applied to titles, abstracts, and keywords, without considering the full text of papers. This may have introduced the risk of missing relevant studies whose quality-related contribution was discussed only in the body of the paper. To mitigate this threat, we complemented the automated search with backward and forward snowballing over the selected primary studies and key secondary studies.Threats to selection bias were mitigated by carefully following established guidelines for systematic mapping studies \cite{a42,a48}. Additionally, the inclusion and exclusion criteria stated in Section 2.1 were formulated by the first two authors and then reviewed and approved by the third and fourth authors to improve clarity and consistency. To reduce data-extraction and classification bias, the extracted data items and category assignments were reviewed by multiple authors, and disagreements were resolved through discussion based on predefined coding criteria.External validity may be limited by the selected publication period, the selected electronic data sources, the English-language restriction, and the focus on the software engineering domain. Therefore, relevant studies from adjacent fields or other indexing services may have been missed. To support transparency and reproducibility, we defined a complete study protocol and provide a replication package containing the search strings, screening decisions, data extraction sheet, and study classifications available at \cite{anonymous_2026_19545171}.

\section{Conclusion}
This paper presented a systematic mapping study on the quality of AI-based software. An initial set of 3628 records retrieved from five electronic databases published between 2020 and 2026, 33 primary studies were selected through systematic screening and snowballing.  The study synthesized the reported challenges, the contributions proposed to address them, and the maturity of the evidence base.

The results show that the most frequently reported challenge concerns limitations in existing quality assessment models, followed by issues related to NFR management, lack of quality-aware development approaches, and inadequate quality assurance.  The study also identified 11 contribution categories, with most research focusing on the quality model enhancement , quality model development, and methodological support. However, the evidence remains dominated by the early-stage and weakly validated research, with limited tool support and industrial validation.

These findings suggest that the field has established a strong conceptual foundation but still requires substantial work to operationalize AI-specific quality attributes and validate proposed approaches in real-word industrial context.  Future research should prioritize empirical validation, stronger tool support, and closer alignment between research, industrial practice, and standardization efforts. 


\begin{credits}
\subsubsection{\ackname} This work has been supported by FAST, the Finnish Software Engineering Doctoral Research Network, funded by the Ministry of Education and Culture, Finland.
\end{credits}
%
%
%
 \bibliographystyle{splncs04}
\bibliography{references}
\end{document}